\documentclass[preprint,prb]{revtex4}
\usepackage{graphicx}
\usepackage{dcolumn}
\usepackage{bm}
\makeatletter
\renewcommand\@biblabel[1]{(#1)}
\makeatother

\begin{document}

\begin{titlepage}

\title{Electronic and Magnetic Properties of Partially-Open Carbon Nanotubes}

\author{Bing Huang$^1$'$^2$, Young-Woo Son$^3$, Gunn Kim$^2$\footnote{E-mail address: gunnkim@phya.snu.ac.kr},
Wenhui Duan$^1$, and Jisoon Ihm$^2$\footnote{E-mail address: jihm@snu.ac.kr}}
\address{$^1$Department of Physics, Tsinghua University, Beijing
100084, People's Republic of China \\ $^2$FPRD and Department of Physics and
Astronomy, Seoul National University, Seoul 151-747, Republic of
Korea \\ $^3$Korea Institute for Advanced Study, Seoul 130-722,
Republic of Korea}
\date{\today}

\begin{abstract}

On the basis of the spin-polarized density functional theory
calculations, we demonstrate that partially-open carbon nanotubes
(CNTs) observed in recent experiments have rich electronic and
magnetic properties which depend on the degree of the opening. A
partially-open armchair CNT is converted from a metal to a
semiconductor, and then to a spin-polarized semiconductor by
increasing the length of the opening on the wall. Spin-polarized
states become increasingly more stable than nonmagnetic states as
the length of the opening is further increased. In addition,
external electric fields or chemical modifications are usable to
control the electronic and magnetic properties of the system. We
show that half-metallicity may be achieved and the spin current may
be controlled by external electric fields or by asymmetric
functionalization of the edges of the opening. Our findings suggest
that partially-open CNTs may offer unique opportunities for the
future development of nanoscale electronics and spintronics.

\end{abstract}

\maketitle

\draft

\vspace{2mm}

\end{titlepage}

\section{Introduction}

Carbon nanotubes (CNTs) are quasi one-dimensional nanostructures
with unique electrical properties that make them ideal candidates
for applications in nanoelectronics\cite{Iijima-1991, Iijima-1993,
Bethune-1993, Tans, Fischer, Javey-2003, Charlier-2007}. Because
bare CNTs alone do not satisfy the requirements for actual
applications, considerable experimental and theoretical efforts have
been directed toward tailoring the electrical and mechanical
properties of CNTs through various methods\cite{Fischer, Javey-2003,
Charlier-2007, SLee-2005, SHKim-2007, MCha-2009, Santos-2009}. Among
them, chemical functionalization/doping and the application of
external eletric/magnetic fields are known as being practically
viable approaches to tailoring the electrical properties of
CNTs\cite{Fischer, Charlier-2007, Sahoo-2005, Fedorov-2005,
Son-2005, Son-2007}. In general, carbon-based systems are promising
candidates for spin-based applications such as spin-qubits and
spintronics\cite{Sahoo-2005, Petta-2005, Son-2006, Nowack-2007,
Awschalom-2007, Tombros-2007, Santos-2009}; they are believed to
have exceptionally long spin coherence times because of the absence
of the nuclear spin in the carbon atom and very weak spin-orbit
interactions. Therefore, in addition to their unique electrical
properties, designing and modulating the spin injection and spin
transport in CNTs have drawn heightened attention.

Recently, some methods have been proposed for lengthwise (\emph{i.
e.}, along the CNT axis) cutting to produce graphene or narrow
graphene nanoribbons (GNRs)\cite{Cano-Marquez-2009, Elias-2009,
Kosynkin-2009, Zenxing-2009, Jiao-2009}; these include etching with
a plasma or an oxidizing agent to unzip the CNT\cite{Kosynkin-2009,
Zenxing-2009, Jiao-2009}. Moreover, the CNT sidewalls can be opened
longitudinally by lithium atoms (or transition metal nanoparticles)
intercalation and followed by exfoliation\cite{Cano-Marquez-2009,
Elias-2009}. It is interesting that the degree of stepwise opening
is controllable by the amount of oxidizing agent or lithium atoms
(or transition metal nanoparticles); partially-open CNTs have been
observed in transmission electron microscopy or scanning electron
microscopy images\cite{Kosynkin-2009, Zenxing-2009,
Cano-Marquez-2009, Elias-2009}. There also exist a few theoretical
studies on the unzipping mechanism of CNTs into GNRs
\cite{Seminario-2009} and the electronic transport in partially
unzipped CNTs\cite{Santos-2009}.

Armchair CNTs are one-dimensional metals with
Dirac-like states crossing the Fermi level, and GNRs have edge
states around the Fermi level which can induce magnetic ground
states\cite{Son-2006}. Since a partially-open CNT is composed of
both a CNT and a curved GNR, it is expected to have more diverse
properties and its electronic properties could be sensitive to the
cutting width (\emph{i.e.}, the circumferential length of cutting)
as well as the cutting length. The edges at the opening may be
chemically active for further chemical modifications as well.

In this article, we systematically investigate physical and chemical
properties of partially-open armchair CNTs using spin-polarized
density functional theory (DFT) calculations. Our results show that
the electronic structures change from a metal to a semiconductor as
the electron is backscattered at the opening (defective region).
When the cutting length increases further, spin-polarization arises
at the zigzag edges of the opening. The spin-polarized states become
increasingly more stable than the nonmagnetic (NM) states following
the increase in the length of cutting. We also demonstrate below
that external electric fields and chemical functionalization are
effective ways of controlling the electronic and magnetic properties
of the partially-open CNTs; they produce still other interesting
properties such as electrical switching and half-metallicity.

\section{Computational Methods and Models}

In the present study, electronic structure calculations were
performed using the Vienna \emph{Ab initio} Simulation Package
(VASP)\cite{VASP} within the framework of the DFT. The projector
augmented wave (PAW) potentials\cite{PAW} and the generalized
gradient approximation with the Perdew-Burk-Ernzerhof (PBE)
functional\cite{PBE} were used to describe the core electrons and
the exchange-correlation energy, respectively. The cutoff energy for
the plane wave basis set was set to 400 eV. Energies were converged
to 10$^{-5}$ eV, and the residual forces on all atoms were converged
to below 0.02 eV/\AA. The supercell for the (8, 8) CNTs of 2.5 nm in
length was considered for typical calculations and the distances
between two adjacent CNTs were maintained at least 10 ~\AA~ to avoid
the interactions between them. Three Monkhost-Pack special
\emph{k}-point meshes yielded $\sim$ 1 meV per atom convergence of
the total energy. The geometrical structures obtained from
optimization processes were used to investigate the electronic and
magnetic behaviors under static external electric fields ({\bf {\it
E}}$_{ext}$); our test calculations for {\bf {\it E}}$_{ext}$ $\leq$
0.5 V/\AA~ on partially-open (8, 8) CNTs showed that the geometry
relaxation from {\bf {\it E}}$_{ext}$ had a negligible effect. {\bf
{\it E}}$_{ext}$ was implemented using a dipole layer in the vacuum,
following widely-used Neugebauer and Scheffler's
method\cite{Neugebauer}.

\begin{figure*}[tbp]
\includegraphics[width=10.0cm]{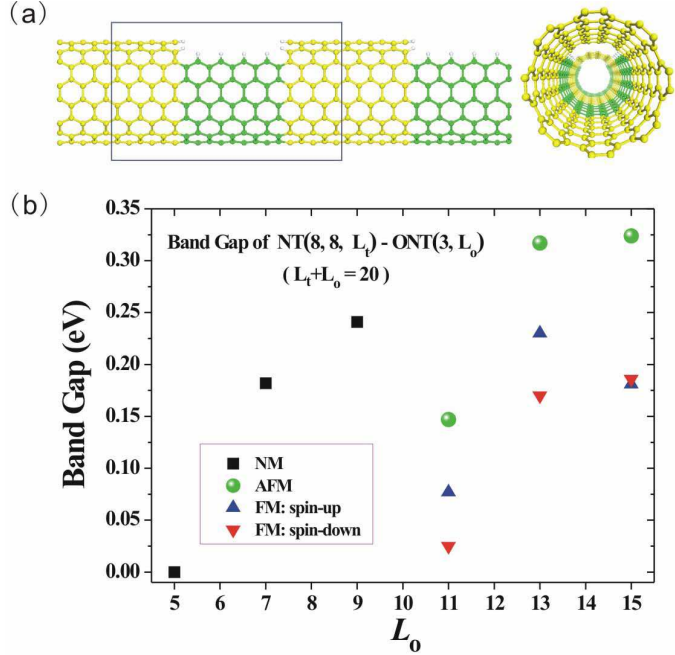}
\caption{(a) Structure of partially-open ($m$, $n$) CNTs (side view
and the perspective view along the tube axis direction). Yellow and
green balls represent carbon atoms on the perfect and the open CNT
parts, respectively. The openings are passivated with hydrogen
atoms, represented as small white balls. The rectangle marks one
unit supercell in the tube axis direction of partially-open CNTs.
The figure shows the structure of NT(8, 8, 11)-ONT(2, 9)
representing a partially-open (8, 8) armchair CNT which has 11 (9)
C-C dimer lines in the perfect CNT part (open part) along the tube
axis direction, and the missing rows in the opening is 2. (b) The
energy band gap as a function of the cutting length $L_{o}$ (for $W$
= 3). The band gaps of nonmagnetic (NM) materials are shown as black
squares ($L_{o} \leq 9$). The band gaps of spin-polarized
semiconductors with antiferromagnetic (AFM) or ferromagnetic (FM)
states are shown as green dots and blue (red) triangles ($L_{o} \geq
11$), respectively.}
\end{figure*}

\section{Results and discussion}

\emph{Geometries, Electronic and Magnetic Properties of
Partially-Open CNTs.} Before discussing our model structures, we
first define a partially-open CNT, as shown in Figure 1a. A
partially-open CNT has two parts: a perfect nanotube part (yellow
color) and an opening (green color). This unit supercell is repeated
periodically in the tube axis direction in the calculations. We
represent the system as NT($m$, $n$, $L_{t}$)-ONT($W$, $L_{o}$)
where $m$ and $n$ correspond to the chiral vector of the CNT,
$L_{t}$ ($L_{o}$) is the length in the tube axis direction of the
perfect (open) CNT in units of carbon columns, and $W$ is the
missing width in units of carbon rows for the opening. For example,
the structure in Figure 1a can be designated as NT(8, 8, 11)-ONT(2,
9).

First, we will show the relation between electronic properties and
the cutting length $L_{o}$ with the cutting width $W$ kept at 3.
Figure 1b summarizes the results for NT(8, 8, $L_{t}$)-ONT(3,
$L_{o}$) with different cutting lengths ($L_{o}$'s) while the total
length of the supercell is kept unchanged ($L_{t}$ + $L_{o}$ = 20).
When the cutting length was small, the overall structural relaxation
was negligible. As the cutting length became longer, the middle part
of the opening was widened to release the compressive stress and
reduce the total energy. This agrees with experimental
observations\cite{Cano-Marquez-2009, Elias-2009, Kosynkin-2009,
Zenxing-2009}. (For optimized structures of NT(8, 8, $L_{t}$)-ONT(3,
$L_{o}$), see Figure S1, Supporting Information.)

\begin{figure*}[tbp]
\includegraphics[width=10.0cm]{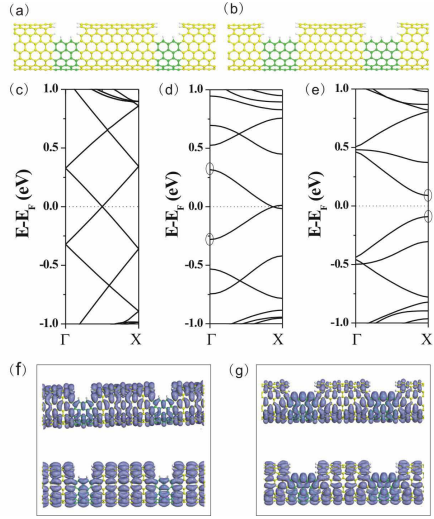}
\caption{Optimized structures (side view) of (a) NT(8, 8, 15)-ONT(3,
5) and (b) NT(8, 8, 13)-ONT(3, 7). (c) Electronic band structure of
the perfect armchair (8, 8) CNT with a supercell size ten times the
primitive unit cell. (d) and (e) are the band structures of NT(8, 8,
15)-ONT(3, 5) and NT(8, 8, 13)-ONT(3, 7), respectively. (f) Side
view of the charge density on the top valence band (lower figure)
and the bottom conduction band (upper figure) at $\Gamma$-point of
the band structure in (d) (marked with circles). (g) Side view of
the charge density on the top valence band (lower figure) and the
bottom conduction band (upper figure) at X-point of the band
structure in (e) (marked with circles). The Fermi level is set to
zero.}
\end{figure*}

For $L_{o}$ $\leq$ 5, the system maintained its metallic character.
The geometric and band structures of NT(8, 8, 15)-ONT(3, 5) are
shown as an example in Figures 2a and 2d. The band structure of the
perfect (8, 8) CNT (the supercell was taken as ten times the
primitive unit cell in order to match the Brillouin zone) is
displayed in Figure 2c for comparison. Obviously, the energy
degeneracies caused by the zone folding at the Brillouin zone
boundary were lifted by the mirror symmetry breaking due to the
existence of the opening (defect); however, no defect states were
generated around the Fermi energy and the system was still metallic.
Charge density analysis (Figure 2f) showed that the charge density
around the Fermi energy was similar to that of a finite armchair
CNT\cite{Rubio-1999} and some standing wave nodes occurred, which
was due to the quantum confinement effects of the opening.

When the cutting length ($L_{o}$) was increased in a stepwise manner
up to 9, ($5 < L_{o} \leq 9$), the system showed NM semiconducting
behaviors. The band gap increased as $L_{o}$ increased, as shown in
Figure 1b. The geometric and band structures of NT(8, 8, 13)-ONT(3,
7) are shown in Figures 2b and 2e. Whereas the perfect (8, 8) CNT
and NT(8, 8, 15)-ONT(3, 5) were metallic, NT(8, 8, 13)-ONT(3, 7) had
a band gap of $\sim$ 0.18 eV. Charge density analysis (Figure 2g)
indicated that the gap opening was due to the breaking of the
degeneracy of the original metallic states around the Fermi level
(E$_{F}$).

For $L_{o} > 9$, the ground state changed from a NM state to a
spin-polarized state. There were two stable spin-polarized states
similar to that of zigzag graphene nanoribbons
(ZGNRs)\cite{Son-2006}, antiferromagnetic (AFM) and ferromagnetic
(FM) state. The AFM configuration had the same spins lining up at
each zigzag edge at the opening but with a relative spin orientation
between two zigzag edges opposite to each other. On the other hand,
the FM configuration had the same spins throughout. Both AFM and FM
states were semiconducting while the NM state (which was unstable
with respect to spin-polarized states) was metallic. The band gap of
the spin-polarized states increased as the cutting length increased,
as plotted in Figure 1b. The AFM state was the ground state for
$L_{o} \geq 11$. For example, the energy of the AFM state was lower
than the NM state by 0.050, 0.121, and 0.138 eV per unitcell for
$L_{o}$ = 11, 13, and 15, respectively. It was also lower in energy
than the FM state by 0.03$-$0.05 eV per unitcell (see Table SI,
Supporting Information). Since our calculations inevitably employed
periodic supercells, a question arises whether our model can
simulate a real system which does not have periodic openings. We
contented ourselves with testing a longer unit cell where the
interaction between two adjacent openings is significantly smaller
than the above one ($L_{t} + L_{o} = 20$). Calculations performed
for a longer unit cell ($L_{t} + L_{o} = 28$) produced essentially
the same results as before (the conversion from a metal to an NM
semiconductor to an AFM semiconductor), suggesting that the periodic
supercell calculation should be a reasonable approximation in the
present case.

\begin{figure*}[tbp]
\includegraphics[width=12.0cm]{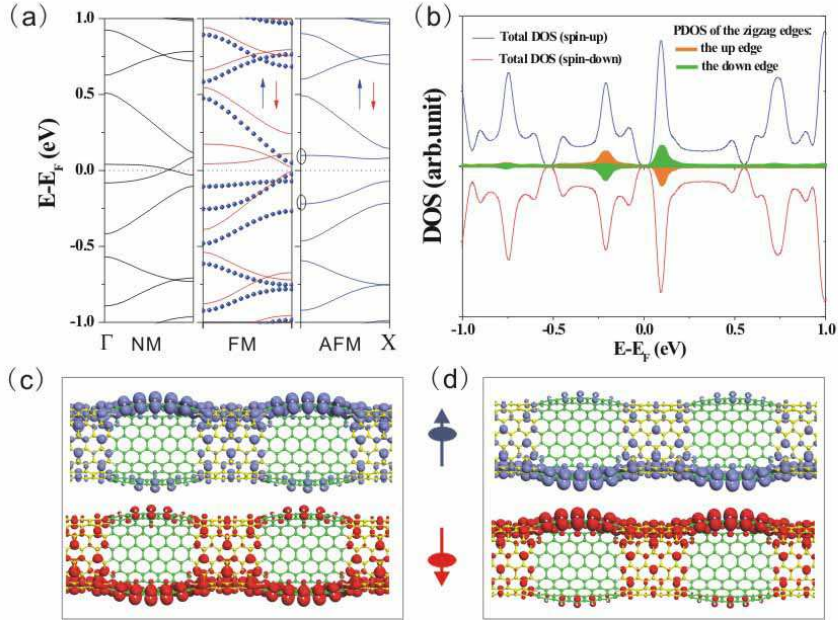}
\caption{Electronic structures and magnetic configurations of partially-open CNTs
(a) Band structures of NT(8, 8, 9)-ONT(3, 11) for the NM,
FM, and AFM states, respectively. Blue and red lines overlap and are
indistinguishable in the AFM case. (b) Density of states (DOS) of
NT(8, 8, 9)-ONT(3, 11) in the AFM ground state. The partial density
of states (PDOS) of zigzag edges at the opening is also plotted as
the green and orange-filled area inside the DOS curves. (c) The
$\Gamma$-point charge density (top view) plots of two spin
degenerate states on the top valence band of AFM state [lower circle
in (a)]. (d) The $\Gamma$-point charge density (top view) plots of
two spin degenerate states on the bottom conduction band of AFM
state [upper circle in (a)]. In these figures, the blue and red
colors represent the spin-up and spin-down states, respectively.}
\end{figure*}

Figure 3 shows the electronic properties of NT(8, 8, 9)-ONT(3, 11)
as a prototype to demonstrate spin-polarized partially-open armchair
CNTs. The NM state was metallic and unstable. Similar to perfect
armchair CNTs, Dirac-like states were crossed at the E$_{F}$, as
shown in the first panel of Figure 3a. Two Dirac-like states at the
crossing point had different slopes, corresponding to different
Fermi velocities. The second panel of Figure 3a displays a
spin-split semiconducting FM state with band gaps of $\sim$ 0.077
and $\sim$ 0.025 eV for the spin-up and spin-down states,
respectively. The magnetic moment of the FM state was close to 2
$\mu_B$. The total energy calculation indicated that the FM state
was not a ground state either; the AFM state was the ground state
with a band gap of $\sim$ 0.147 eV (or even larger gap for longer
$L_{o}$). Spin distribution analysis (Figure S2, Supporting
Information) showed that the spin polarization was mainly localized
at the zigzag edges of the opening and that the spin moment at the
center of the opening was largest ($\sim$ 0.23 $\mu_B$ per atom).
Small spin moments ($\sim$ 0.05 $\mu_B$ per atom) were induced in
the perfect CNT part. The states of the open zigzag edges were
mainly located around the top valence band and the bottom conduction
band; these are shown in the partial density of states (PDOS)
(Figure 3b) and charge density plots (Figures 3c and 3d). The two
spin states were located separately at opposite sides of the open
zigzag edges, similar to those of ZGNRs\cite{Son-2006}. Our previous
work showed that the spin-polarization of ZGNRs could be understood
in terms of the Stoner magnetism for \emph{sp} electrons (instead of
conventional \emph{d} electrons) occupying a very narrow edge band
(thus a huge density of states) to induce instability toward
spin-band splitting\cite{Bing-PRB}. Unlike ZGNRs, the NM state here
only had Dirac-like bands crossing at the E$_{F}$. However, due to
the opening, the DOS of the NM state near E$_{F}$ increased
(\emph{i. e.}, the bands were flattened near E$_{F}$), as the
cutting length $L_{o}$ increased from 11 to 15, as shown in Figure
4. This presumably led to instability of the NM state and stabilized
the spin-polarized state, analogous to ZGNRs.

\begin{figure*}[tbp]
\includegraphics[width=10.0cm]{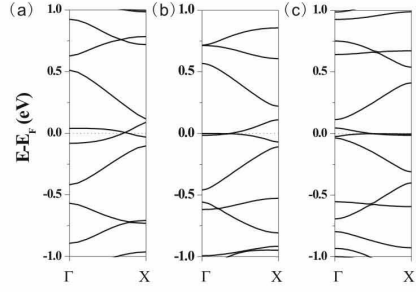}
\caption{The nonmagnetic band structures of (a) NT(8, 8, 9)-ONT(3,
11), (b) NT(8, 8, 7)-ONT(3, 13) and (c) NT(8, 8, 5)-ONT(3, 15).}
\end{figure*}

\begin{table}
\caption{\label{tab:table2} Energy band gap (in eV) as a function of
the cutting width $W$ for three different cutting length.}
\begin{ruledtabular}
\begin{tabular}{ccccc}
$W$ & 2 & 3 & 4 &
 \\[2pt]
\hline
\ $L_{o}$ = 5 (NM) & 0  & 0 & 0.061  \\[2pt]
\ $L_{o}$ = 9 (NM) & 0.160  & 0.241 & 0.326 \\[2pt]
\ $L_{o}$ = 11 (AFM) & 0.112  & 0.147 & 0.195 \\[2pt]
\end{tabular}
\end{ruledtabular}
\end{table}

We also show in Table I the relation between the electronic
structure and the cutting width ($W$) for three different cutting
lengths. For a short cutting length ($L_{o}$ = 5), when $W$
increased, a small band gap opened up because of the increased
backscattering from the opening. For the middle cutting length
($L_{o}$ = 9), the band gap increased significantly as $W$
increased. For the spin-polarized one ($L_{o}$ = 11), the band gap
of the AFM ground state also became larger as $W$ increased, while
the stability of the AFM states (the energy difference between the
AFM and NM states) did not change much (see Table SII, Supporting
Information).

In order to investigate the curvature effect on electronic and
magnetic properties of partially-open CNTs, we also calculated the
electronic structures of (6, 6) and (10, 10) CNTs. Here, all the
CNTs had the same supercell length along the axial direction.
Similar to the (8, 8) CNT, the (6, 6) and (10, 10) CNTs showed the
conversion from a metal to a semiconductor to a spin-polarized
semiconductor by increasing the cutting length $L_{o}$. The relation
between the band gap and the diameter of the CNT is listed in Table
II. Two different cuttings corresponding to a nonmagnetic system
($W$ = 3, $L_{o}$ = 7) and a spin-polarized structure ($W$ = 2,
$L_{o}$ = 11) were considered. For the nonmagnetic system ($W$ = 3,
$L_{o}$ = 7), the band gap decreased remarkablely from 0.337 eV to
0.099 eV, as the diameter of CNT increased. For the spin-polarized
system ($W$ = 2, $L_{o}$ = 11), the AFM band gaps (for the ground
state) decreased little (from 0.117 eV to 0.099 eV) as the diameter
of CNT increased, demonstrating that the AFM band gap was
insensitive to the diameter of the CNT.

\begin{table}
\caption{\label{tab:table2} Energy band gap (eV) as a function of
the diameter of CNT for two different cuttings. ($L_{t}$ + $L_{o}$ =
20)}
\begin{ruledtabular}
\begin{tabular}{ccccc}
($W$, $L_{o}$) & (6, 6) CNT & (8, 8) CNT & (10, 10) CNT &
 \\[2pt]
\hline
\ (3, 7) (NM) & 0.337  & 0.182 & 0.099  \\[2pt]
\ (2, 11) (AFM) & 0.117  & 0.112 & 0.099 \\[2pt]
\end{tabular}
\end{ruledtabular}
\end{table}

\begin{figure*}[tbp]
\includegraphics[width=12.0cm]{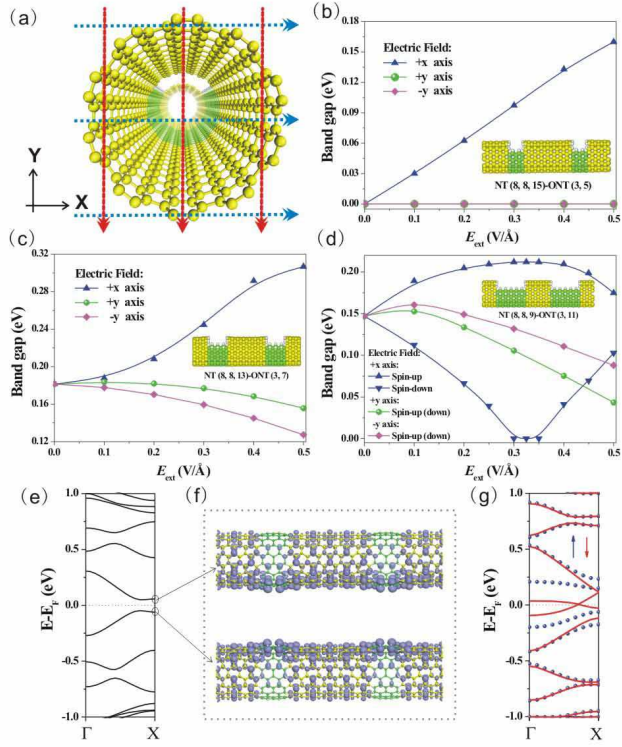}
\caption{Electronic properties of partially-open nanotubes with {\bf
{\it E}}$_{ext}$. (a) The model of partially-open armchair CNTs with
{\bf {\it E}}$_{ext}$. The band gaps as a function of {\bf {\it
E}}$_{ext}$ are plotted for (b) NT(8, 8, 15)-ONT(3, 5), (c) NT(8, 8,
13)-ONT(3, 7), and (d) NT(8, 8, 9)-ONT(3, 11), respectively; the
corresponding atomic structures (side view) are also shown as an
inset in each case. (e) The band structure of NT(8, 8, 15)-ONT(3, 5)
with {\bf {\it E}}$_{ext}$ = 0.3 V/\AA~ in the $+x$ direction. (f)
Top view of the charge density on the top valence band (lower
figure) and the bottom conduction band (upper figure) at the X-point
of the band structure in (e) (marked with circles). (g)
Spin-polarized band structure of NT(8, 8, 9)-ONT(3, 11) with {\bf
{\it E}}$_{ext}$ = 0.3 V/\AA~ in the $+x$ direction.}
\end{figure*}

\emph{Homogeneous External Electric Fields for Modulating Properties
of Partially-Open CNTs.} The response of CNTs to an external
electric field ({\bf {\it E}}$_{ext}$) is an interesting subject for
future applications\cite{Sahoo-2005, Fedorov-2005}.  The three
directions of {\bf {\it E}}$_{ext}$ ($+x$, $+y$, $-y$), which were
all perpendicular to the tube axis direction, $z$, were considered.
It is noted that $+x$ is equivalent to $-x$ while $+y$ is not
equivalent to $-y$ in the partially-open CNTs. Before the
calculations of our system, some tests were done on the ZGNRs, using
the {\bf {\it E}}$_{ext}$ method and the GGA approximation with the
PBE functionals. Our tests showed that {\bf {\it E}}$_{ext}$ effect
obtained from the GGA was significantly larger than that from the
local density approximation (LDA); this was consistent with previous
observations\cite{Hod-2007, Kan-2007}. For example, the critical
value of {\bf {\it E}}$_{ext}$ for half-metallicity in 16-ZGNR in
our GGA calculations was $\sim$ 0.55 V/\AA, which was much larger
than that from LDA calculations ($\sim$ 0.20 V/\AA), and the
difference agreed with previous works\cite{Hod-2007, Kan-2007}.

Figure 5 briefly summarizes the results of partially-open armchair
CNTs affected by {\bf {\it E}}$_{ext}$. Figure 5a shows the geometry
and the directions of the {\bf {\it E}}$_{ext}$, and Figures 5b-5d
show the band gap behaviors for different cases. Our calculations on
perfect armchair CNTs demonstrated that the band structure
experienced little change in response to {\bf {\it E}}$_{ext}$; this
was due to the strong screening behavior, which agreed well with
previous results\cite{Son-2005, James-2002, Yan-2003}. For metallic
NT(8, 8, 15)-ONT(3, 5) shown as the inset in Figure 5b, the system
was still metallic when {\bf {\it E}}$_{ext}$ was in the $+y$ or
$-y$ direction, as in the case of the perfect armchair CNT. However,
when {\bf {\it E}}$_{ext}$ was applied in the $+x$ direction, the
band gap opened up and drastically increased with the increase in
the field strength. The band structure of NT(8, 8, 15)-ONT(3, 5) at
0.3 V/\AA~ in the $+x$ direction is shown in Figure 5e. Compared
with the unperturbed band structure (Figure 2d), the energy degeneracies
were broken and a band gap appeared at the Fermi
energy. The charge density of the top valence band and the bottom
conduction band at the X-point are shown in Figure 5f. With {\bf
{\it E}}$_{ext}$ in the $+x$ direction, the cylindrical symmetry of
the charge density was broken, the top valence band around the
X-point was localized mainly on the upper part of the tube, and the
bottom conduction band around the X-point was localized mainly on
the lower part of the tube. Therefore, the origin of the band gap
opening was the Stark effect. For perfect armchair CNTs, extremely
large {\bf {\it E}}$_{ext}$ ($> 1.5$ V/\AA) was required for the
band gap opening \cite{Son-2005, James-2002, Yan-2003}. NT(8, 8,
15)-ONT(3, 5) was now polarized with an induced dipole moment
created by {\bf {\it E}}$_{ext}$ (Figure S3, Supporting
Information). The charge density at the top valence band and the
bottom conduction band changed little if {\bf {\it E}}$_{ext}$ was
in the $+y$ or $-y$ direction. Thus, band structures showed little
perturbation. In the case of semiconducting NT(8, 8, 13)-ONT(3, 7),
shown as the inset of Figure 5c, the band gap increased to 0.30 eV
when {\bf {\it E}}$_{ext}$ in the x direction reached 0.50 V/\AA.
The band gap decreased by only $\sim$ 0.03 eV ($\sim$ 0.06 eV) at
0.50 V/\AA~ in the $+y$ ($-y$) direction, and the redistributions of
electrons at the top valence band and the bottom conduction band was
not as pronounced as for {\bf {\it E}}$_{ext}$ in the $+x$
direction.

For NT(8, 8, 9)-ONT(3, 11) shown as an inset of Figure 5d, since the
oppositely oriented spin states were located at opposite sides of
the zigzag edge of the opening (Figure 3), the effect of {\bf {\it
E}}$_{ext}$ on the spin states in the $+x$ direction on them should
be opposite. This led to the occupied spin-down states moving up and
unoccupied spin-down states moving down, making the distance in
energy closer, as shown in Figure 5d. The spin-up states behaved
oppositely, enlarging the energy gap. This phenomenon was similar to
the ZGNRs under transverse {\bf {\it E}}$_{ext}$\cite{Son-2006}. The
critical {\bf {\it E}}$_{ext}$ strength for half-metallic states
(closing the band gap of the spin-down state) was around 0.3 V/\AA.
The corresponding band structure is shown in Figure 5g.
Interestingly, the electronic structure of spin-down states around
the Fermi energy was Dirac-like, unlike in ZGNRs. When {\bf {\it
E}}$_{ext}$ was applied in the $+y$ or $-y$ direction, the band gaps
of both spin states increased a little initially but decreased
significantly under a stronger external field. Similar to the
metallic open CNTs (Figure 5b) or the NM semiconductors (Figure 5c),
the redistributions of the charge densities at the top valence band
and the bottom conduction band was not pronounced for {\bf {\it
E}}$_{ext}$ in the $+y$ or $-y$ direction. However, unlike in
Figures 5b and 5c, {\bf {\it E}}$_{ext}$ in the $\pm y$ direction
influenced the band gap more because the partially localized states
around the E$_{F}$ were substantially influenced by {\bf {\it
E}}$_{ext}$ and thereby changing the band gap\cite{Son-2005,
Son-2006, Son-2007, Tien-2005}. Based on the behavior of spin states
shown in Figures 5d and 5g, we propose that the partially-open CNT
with {\bf {\it E}}$_{ext}$ may be used as a spin source to inject
spin currents into the perfect CNT which is seamlessly connected to
it.

We can explain the modification of band structures of partially-open
CNTs in terms of the symmetry breaking as follows. For a perfect
armchair CNT, the cylindrical symmetry and the mirror reflection
symmetry are preserved. The $\pi$-bonding and $\pi$-antibonding
($\pi^*$) states are degenerate eigenstates crossing at the Fermi
energy in this case. The wave functions in the $\pi$ band do not
have phase variation along the circumference of the tube whereas the
signs of the wave functions in the $\pi^*$ band alternate
rapidly\cite{GKim-2005}. However, the symmetry of our model
structures becomes lower because of the openings. If the
origin of the coordinate is set on the axis of the nanotube, our
model structures retain the mirror reflection symmetry with respect
to the $x$ = 0 plane in the absence of the external fields. In
contrast, they have no mirror reflection symmetry with respect to
the $y$ = 0 plane. When {\bf {\it E}}$_{ext}$ is applied in the $\pm
x$ direction, the mirror reflection symmetry with respect to the $x$
= 0 plane is broken as well. In Figure 5b, $\pi$ and $\pi^*$ states
are no longer eigenstates in this situation and they become
hybridized. Now the degeneracy is lifted and the band gap opens up
rapidly as {\bf {\it E}}$_{ext}$ increases. For {\bf {\it
E}}$_{ext}$ in the $\pm y$ direction, the mirror symmetry remains
for the $x$ = 0 plane and metallic partially-open CNTs maintain
their metallic characters shown in Figure 5b. Since the screening
effects of semiconducting partially-open CNTs are weak, however,
electronic structures of these CNTs do change a little bit even by
{\bf {\it E}}$_{ext}$ in the $\pm y$ direction as shown in Figure
5c.

\emph{Chemical Functionalization for Modifying Properties of
Partially-Open CNTs.} Finally, we turn to chemical functionalization
of CNTs. Chemical functionalization is an effective way to tailor
the electronic and transport properties of CNTs\cite{Fischer,
Charlier-2007}. In an oxidative process, oxygen-containing
functionalities such as carbonyl (C=O), carboxyl (COOH), and
hydroxyl (OH) groups were observed to exist at the edge and on the
surface of CNTs\cite{Kosynkin-2009, Zenxing-2009, Liu-1998,
Dujardin-1998, Zhao-2002}. As in the previous cases of the hydrogen
passivation, the OH groups passivated the dangling $sp^2$ $\sigma$
bonds at the open edges. Carbon $\pi$ orbitals were affected by OH
or H through the shift of the effective potential at the edge carbon
atoms. When the edges in the opening were passivated with OH groups
instead of H atoms, the overall shapes of the band structures did
not change much (Figure S4, Supporting Information) but the band gap
dramatically decreased. For example, the band gaps of NT(8, 8,
13)-ONT(3, 7) and NT(8, 8, 9)-ONT(3, 11) decreased by $\sim$ 50 \%
from 0.182 and 0.147 eV to 0.095 and 0.080 eV, respectively, shown
in Figure S4 in Supporting Information. According to our
calculations, the bond length between the O atom in the OH group and
the edge carbon atom was $\sim$ 1.36 \AA.

\begin{figure*}[tbp]
\includegraphics[width=12.0cm]{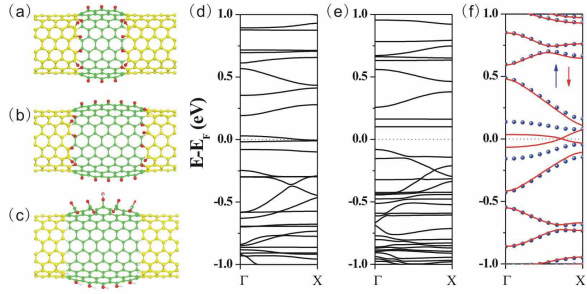}
\caption{Geometric and electronic structures of partially-open
armchair CNTs with open edges modified with chemical groups: (a)
NT(8, 8, 13)-ONT(3, 7) with O-termination; (b) NT(8, 8, 9)-ONT(3,
11) with O-termination; (c) NT(8, 8, 9)-ONT(3, 11) with two zigzag
edges terminated with COOH and OH groups, respectively. (d), (e),
and (f) correspond to electronic structures for (a), (b), and (c),
respectively.}
\end{figure*}

Compared to H- or OH-termination, O-termination strongly modified
the overall electronic structures of partially-open CNTs. In this
case, each O atom was bonded to an edge carbon atom at the opening
and its bond length was $\sim$ 1.26 \AA. In comparison, a carbon
dioxide (CO$_2$) molecule has a C=O bond of $\sim$ 1.2 \AA. We chose
O-terminated NT(8, 8, 13)-ONT(3, 7) and NT(8, 8, 9)-ONT(3, 11) to
represent the spin-unpolarized and spin-polarized partially-open
armchair CNTs, respectively. The optimized structures are depicted
in Figures 6a and 6b.  For NT(8, 8, 13)-ONT(3,7), some localized
states crossed at the Fermi energy, as shown in Figure 6d, and the
charge density showed that edge O atoms and neighboring carbon atoms
made major contributions to these states. The conductance of NT(8,
8, 13)-ONT(3, 7) was greatly affected by those localized states. For
NT(8, 8, 9)-ONT(3, 11), the spin-polarization was suppressed and the
ground state turned to the NM state, shown in Figure 6e, which was
mainly due to the double bonds between C and O, destroying the edge
states of the zigzag edges. Besides, some flat band states localized
around the O atoms and the openings were produced below the bottom
of the conduction band. Our results clearly showed that the oxygen
should be avoided for spintronics applications of partially-open
CNTs.

We also studied the electronic properties of partially-open CNTs
with asymmetric edge functionalization  as demonstrated for the
ZGNRs\cite{Kan-2008}. We chose the COOH groups and the OH groups to
terminate the two open zigzag edges of the spin-polarized NT(8, 8,
9)-ONT(3, 11) structure, respectively, as shown in Figure 6c. The
bond length between the carbon atom in the COOH group and the edge
carbon atom at the opening was $\sim$ 1.47 \AA. The COOH group acted
as an electron-acceptor whereas the OH group was effectively an
electron donor. The asymmetric edge termination with COOH and OH
groups resulted in a large potential difference between the two
edges, and the effect was similar to applying electric fields in the
$x$ direction as studied before (Figures 5d and 5g). Figure 6f shows
the spin splitting effect and the system became a half-metal as
expected from the analogy to applied electric fields. Here, the
half-metallicity was permanently established without external
electric fields. Thus, we can conclude that the chemical
modification is an efficient method of tailoring spin properties of
our proposed system.

\section{Summary}

Using the spin-polarized density functional theory, we demonstrated
that partially-open armchair CNTs had unusual electronic properties
depending on the degree of opening. A partially-open armchair CNT
was converted from a metal to a semiconductor to a spin-polarized
semiconductor by increasing the opening length. The spin-polarized
states became more stable than nonmagnetic states with increase in
the opening length. In addition, external electric fields and
chemical functionalization were shown to be utilizable for
controlling the electric and magnetic properties, as well as
producing still other interesting properties such as electrical
switching and half-metallicity.

\section*{Acknowledgments}
The authors acknowledge the supports by the Core Competence
Enhancement Program (2E21040) of KIST through the Hybrid
Computational Science Laboratory, the SRC program (Center for
Nanotubes and Nanostructured Composites) of MEST/KOSEF, the Korean
Government MOEHRD, Basic Research Fund No. KRF-2006-341-C000015
(J.I.), the second BK21 program (G.K.), and KOSEF grant (Quantum
Metamaterials research center, No. R11-2008-053-01002-0) funded by
the MEST (Y.-W.S.). This work was also supported by the A3 Foresight
Program of KOSEF-NSFC-JSPS, the Ministry of Science and Technology
of China (Grant Nos. 2006CB605105, 2006CB0L0601), and the National
Natural Science Foundation of China (W.D. and B.H.). The
computations were performed through the support of KISTI in Korea.

\section*{Supporting Information Available:} This material is
available free of charge via the Internet at http://pubs.acs.org.



\end{document}